\documentclass[twocolumn,showpacs,preprintnumbers,amsmath,amssymb]{revtex4}
\usepackage{graphicx}% Include figure files
\usepackage{dcolumn}% Align table columns on decimal point
\usepackage{bm}% bold math

\usepackage{amssymb}

\begin{document}

% \preprint{APS/123-QED}

\title{Constraints on the IR behavior of the gluon propagator
        in Yang-Mills theories}

\author{A. Cucchieri}
\affiliation{Instituto de F\'{\i}sica de S\~ao Carlos, Universidade de
             S\~ao Paulo, \\ Caixa Postal 369, 13560-970 S\~ao Carlos, SP, Brazil}%
\author{T. Mendes\footnote{Permanent address: 
             Instituto de F\'{\i}sica de S\~ao Carlos, Universidade de
             S\~ao  Paulo, C.P.\ 369, 13560-970 S\~ao Carlos, SP, Brazil.}%
        }
\affiliation{DESY--Zeuthen, Platanenallee 6, 15738 Zeuthen, Germany}%

\date{\today}% It is always \today, today,
             %  but any date may be explicitly specified

\begin{abstract}
We present rigorous upper and lower bounds for the zero-momentum gluon
propagator $D(0)$ of Yang-Mills theories in
terms of the average value of the gluon field. This 
allows us to perform a controlled extrapolation of lattice data to 
infinite volume, showing that the infrared limit of the Landau-gauge 
gluon propagator in $SU(2)$ gauge theory is finite and nonzero
in three and in four space-time dimensions. In the two-dimensional
case we find $D(0) = 0$, in agreement with Ref.\ \cite{Maas:2007uv}. 
We suggest an explanation for these results. We note that our discussion
is general, although we only apply our analysis to pure gauge theory in 
Landau gauge. Simulations have
been performed on the IBM supercomputer at the University of S\~ao Paulo.
\end{abstract}

% PACS, the Physics and Astronomy
                                % Classification Scheme.
%\keywords{Suggested keywords}%Use showkeys class option if keyword
                              %display desired
\maketitle

%%%%%%%%%%%%%%%%%%%%%%%%%%%%%%%%%%%%%%%%%%%%%%%%%%%%%%%%%%%%%%%%%%%%%%%%%%%%%%%%%%%%%%%%

\section{Introduction}

Color confinement is a basic feature of hadron physics that still lacks a clear
theoretical understanding. Among several explanations suggested in the
literature (see \cite{Alkofer:2006fu} for a recent review), the so-called
Landau-gauge Gribov-Zwanziger scenario \cite{Gribov,vanishing} relates
gluon confinement to the infrared (IR) suppression of the gluon propagator 
$D(p^2)$, whereas quark confinement is related to the IR enhancement of the 
ghost propagator $G(p^2)$. 
This scenario is supported by several studies using functional methods
\cite{Fischer:2006ub}. In particular, these studies \cite{SD1,SD2,Lerche:2002ep}
predict, for small momenta, a gluon propagator $D(p^2) \propto p^{2(a_D - 1)}$ and
a ghost propagator $G(p^2) \propto 1/ p^{2(1 + a_G)}$. The IR exponents
$a_D$ and $a_G$ should satisfy the relation
$a_D  =  2 a_G + (4 - d)/2 \, $,
where $d$ is the space-time dimension and $a_G$ should have a value in the
interval $[1/2, 1]$. Clearly, if $a_D > 1$ one has $D(0) = 0$, implying maximal
violation of reflection positivity \cite{vanishing}. In the four-dimensional case
one finds \cite{SD2,Lerche:2002ep} $a_G \approx 0.59$ and $a_D = 2 a_G$. Similar
power behaviors have also been obtained for the various vertex functions of SU($N_c$)
Yang-Mills theories \cite{Lerche:2002ep,Fischer:2006vf}. As a consequence, the running
coupling constants from the ghost-gluon, three-gluon and four-gluon vertices
are all finite at zero momentum, displaying a universal (qualitative) behavior
\cite{Fischer:2006ub}. Let us note that a key ingredient of these results is
the non-renormalizability of the ghost-gluon vertex, i.e.\ ${\widetilde Z}_1(p^2) = 1$,
which has been verified at the nonperturbative level \cite{vertex}
using lattice Monte Carlo simulations.

One should stress, however, that different IR behaviors for the Landau gluon and 
ghost propagators have also been proposed in the literature. For example in Ref.\
\cite{Aguilar:2004sw} the authors find that $D(0)$ is finite and nonzero and that
$a_G \approx 0$, with a gluon propagator characterized by a dynamically generated mass.
Similar results are obtained in \cite{Frasca:2007uz}. On the other hand, in Refs.\
\cite{Boucaud}, using Ward-Slavnov-Taylor identities, the authors conclude
that $D(p^2)$ should be (probably very weakly) divergent at small momenta and that
$a_G = 0$. Recently, Chernodub and Zakharov \cite{Chernodub:2007rn} obtained the
relation $ 2 a_D + a_G = 1$ for the 4d IR exponents of gluon and ghost propagators,
by considering the contribution of these propagators to thermodynamic quantities
of the system, such as pressure and energy density. This result, together with
the previous relation between $a_D$ and $a_G$,
implies $a_D = 2/5$ and $a_G = 1/5$, i.e.\ the ghost
propagator blows up faster than $p^{-2}$ at small momenta, while the gluon
propagator diverges as $p^{-6/5}$. Very recently, in Ref.\ \cite{silvio}, it was shown
that using the Gribov-Zwanziger approach one can also obtain a finite $D(0)$
gluon propagator and $a_G = 0$.
 Finally, phenomenological tests \cite{Natale:2006nv}
seem to favor a finite and nonzero $D(0)$.

Numerical studies using Monte Carlo simulations suggest that the
gluon propagator is finite at zero momentum \cite{gluonghost,gluon,
Cucchieri:2006xi,Maas:2007uv,3d} and that the ghost propagator
\cite{gluonghost,ghost,Maas:2007uv} is enhanced
when compared to the tree-level behavior $p^{-2}$. Moreover, in 2d and in 3d
\cite{Maas:2007uv,3d} the gluon propagator $D(p^2)$
shows a maximum value for $p$ of a few hundred MeV and
decreases as $p$ goes to 0. On the other hand, in 4d, even using lattices with a lattice
side of about $10 \,\mbox{fm}$, one does not see a gluon propagator decreasing
at small momenta \cite{Cucchieri:2006xi}. It has been argued that an
IR decreasing gluon propagator can be obtained numerically only when
simulations are done on large enough lattice sizes \cite{manifold}.
However, from recent studies in 4d using very large lattice volumes
\cite{Cucchieri:2007md,Bogolubsky:2007ud,Sternbeck:2007ug},
one sees that $D(p^2)$ either displays a plateau for momenta
$p \lesssim 100$ MeV or gets slightly suppressed at small momenta.
Let us note that one of the main problems of the numerical studies
of the gluon propagator
is the lack of a simple way of extrapolating the data to infinite volume.

Here we discuss the behavior of the gluon propagator at zero momentum. We
show that, instead of studying $D(0)$ directly, it is more convenient
to consider the quantity
\begin{equation}
{M}(0) \; = \; \frac{1}{d (N_c^2 - 1)}
  \sum_{\mu, b} | A^b_{\mu}(0) | \; .
\label{eq:Aabs}
\end{equation}
In a spin system this would be equivalent to studying the average absolute value 
of the components of the magnetization instead of the susceptibility, which is
of course a much noisier quantity. 
(Note that, by symmetry, the field components will average to zero if no
absolute value is taken.)
In order to relate ${M}(0)$ to $D(0)$
we derive in Section \ref{sec:ineq} rigorous lower and upper bounds for
$D(0)$, which are expressed in terms of ${M}(0)$. Numerical data are obtained
from extensive simulations in two, three and in four dimensions,
for the pure SU(2) case, using very large lattices in the scaling region.
We show in Section \ref{sec:results} that using these bounds 
for $D(0)$ and with present lattice sizes we
have clear control over the extrapolation of the data to the
infinite-volume limit. In the same section we suggest a possible
explanation of the results obtained. Finally, in Section \ref{sec:conclusions}
we present our conclusions. We note that our discussion concerning the bounds for
the gluon propagator is general, although we only consider here the Landau-gauge 
propagator and pure $SU(2)$ gauge theory.
Note also that recent studies \cite{Cucchieri:2007zm,
Sternbeck:2007ug} have verified the analytic prediction that Landau-gauge gluon 
and ghost propagators in SU(2) and in SU(3) are rather similar. 
Thus, we expect that the analysis presented here
should apply also to the SU(3) case.

%%%%%%%%%%%%%%%%%%%%%%%%%%%%%%%%%%%%%%%%%%%%%%%%%%%%%%%%%%%%%%%%%%%%%%%%%%%%%%%%%%%%%%%%

\section{Lower and upper bounds for $D(0)$}
\label{sec:ineq}

As said in the Introduction, interesting lower and upper bounds for the gluon
propagator at zero momentum $D(0)$ can be obtained by considering the quantity
${M}(0)$ defined in Eq.\ (\ref{eq:Aabs}), i.e.\ the average of the absolute
value of the components of the gluon field at zero momentum. These components
are given by
\begin{equation}
A^b_{\mu}(0) \; = \; \frac{1}{V} \sum_x A^b_{\mu}(x) \; .
\label{eq:A0field}
\end{equation}
In Refs.\ \cite{vanishing} it was shown that in Landau and in Coulomb gauge
the quantity ${M}(0)$ should go to zero at least as fast as $1/N$ in the
infinite-volume limit, where $N$ is the number of lattice points per direction.
This result is simply a consequence of the positivity of the Faddeev-Popov
matrix, i.e.\ it applies to gauge-fixed configurations that belong to the interior
of the first Gribov region.

In order to find the lower and upper bounds for $D(0)$ lets us consider the
inequality
\begin{equation}
\left( \frac{1}{m} \sum_{i=1}^m x_i \right)^2 \, \leq \;
       \frac{1}{m} \sum_{i=1}^m x_i^2 \; ,
\label{eq:ineqCBS}
\end{equation}
where $ \vec{x} $ is a vector with $m$ components $ x_i $.
This result simply says that the square of the average of an observable
is smaller than or equal to the average of the square of this quantity and 
is equivalent to the inequality
\begin{equation} 
\frac{1}{m} \sum_{i=1}^m \left( x_i - {\overline x} \right)^2 \, \geq \, 0
               \; , \qquad \;
                  {\overline x} \, = \, \frac{1}{m} \sum_{i=1}^m x_i \; .
\end{equation}
Note that expression (\ref{eq:ineqCBS}) becomes an equality
when $x_i = \mbox{constant}$. We now apply (\ref{eq:ineqCBS})
to the vector with $ m = d (N_c^2 - 1)$ components
$ \langle | A^b_{\mu}(0) | \rangle $. This yields
\begin{equation}
{\langle {M}(0) \rangle}^2 \, \leq \;
       \frac{1}{d (N_c^2 - 1)} \sum_{\mu, b} \langle \, | A^b_{\mu}(0) | \, \rangle^2 \; .
\label{eq:ineq1}
\end{equation}
Then, we can apply the same inequality to the Monte Carlo
estimate of the average value
\begin{equation}
\langle \, | A^b_{\mu}(0) | \, \rangle \; = \; \frac{1}{n} \sum_c | A^b_{\mu, c}(0) | \; ,
\end{equation}
where $n$ is the number of configurations. In this case we obtain
\begin{equation}
\langle \, | A^b_{\mu}(0) | \, \rangle^2 \, \leq \;
       \langle \, | A^b_{\mu}(0) |^2 \, \rangle \; .
\label{eq:ineq2}
\end{equation}
Thus, by recalling that
\begin{equation}
D(0) \; = \; \frac{V}{d (N_c^2 - 1)} \sum_{\mu, b} \langle | A^b_{\mu}(0) |^2 \rangle \; ,
\end{equation}
and using Eqs.\ (\ref{eq:ineq1}) and (\ref{eq:ineq2}) we find
\begin{equation}
V \, {\langle {M}(0) \rangle}^2 \, \leq \; D(0) \;. \label{eq:ineq1D0}
\end{equation}
At the same time we can write the inequality
\begin{equation}
\langle \, \sum_{\mu, b} | A^b_{\mu}(0) |^2 \, \rangle
\, \leq \; \langle \Bigl\{ \, \sum_{\mu, b} | A^b_{\mu}(0) | \, \Bigr\}^2 \rangle \; .
\end{equation}
This implies
\begin{equation}
D(0) \, \leq \; V d (N_c^2 - 1) \, \langle {{M}(0)}^2 \rangle \; . 
\label{eq:ineq2D0}
\end{equation}
Thus, if ${M}(0)$ goes to zero as $V^{-\alpha}$ we find that $D(0) \to 0$,
$0 < D(0) < +\infty$ or $D(0) \to +\infty$ respectively if the exponent $\alpha$ is
larger than, equal to or smaller than 1/2.
Finally, let us note that the inequalities (\ref{eq:ineq1D0}) and (\ref{eq:ineq2D0}) can
be immediately extended to the case $D(p^2)$ with $p \neq 0$.

%%%%%%%%%%%%%%%%%%%%%%%%%%%%%%%%%%%%%%%%%%%%%%%%%%%%%%%%%%%%%%%%%%%%%%%%%%%%%%%%%%%%%%%%

\section{Results}
\label{sec:results}

We have considered several lattice volumes in 2d (at $\beta = 10$, up to
a lattice volume $V = 320^2$)
in 3d (at $\beta = 3$, up to $V = 320^3$) and in 4d (at $\beta = 2.2$,
up to $V = 128^4$).
Details of the simulations will be presented elsewhere \cite{prepar}. 
We set the lattice spacing $a$
by considering the input value
$\sigma^{1/2} = 0.44$ GeV, which is a typical value for this quantity in the
4d SU(3) case.
Note that the lattice volumes $320^2$ at $\beta = 10$, $320^3$ at $\beta = 3$
and $128^4$ at $\beta = 2.2$ correspond (respectively) to $V \approx (70 \,\mbox{fm})^2$,
$V \approx (85 \,\mbox{fm})^3$ and to $V \approx (27 \,\mbox{fm})^4$.
Simulations in 2d have been done on a PC cluster at the IFSC--USP (with
4 PIV 2.8GHz and 4 PIV 3.0GHz). 
Simulations in 3d and in 4d have been done in the 4.5 Tflops IBM supercomputer at USP
\cite{lcca}. The total CPU-time was equivalent to about 5.7 days (in 3d) and
25.9 days (in 4d) on the whole machine.

%%%%%%%%%%%%%%%%%%%%%%%%%%%%%%%%%%%%%%%%%%%%%%%%%%%%%%%%%%%%%%%%%%%%%%%%%%%%%%%%%%%%%%%%

\begin{figure}
\includegraphics[scale=0.32]{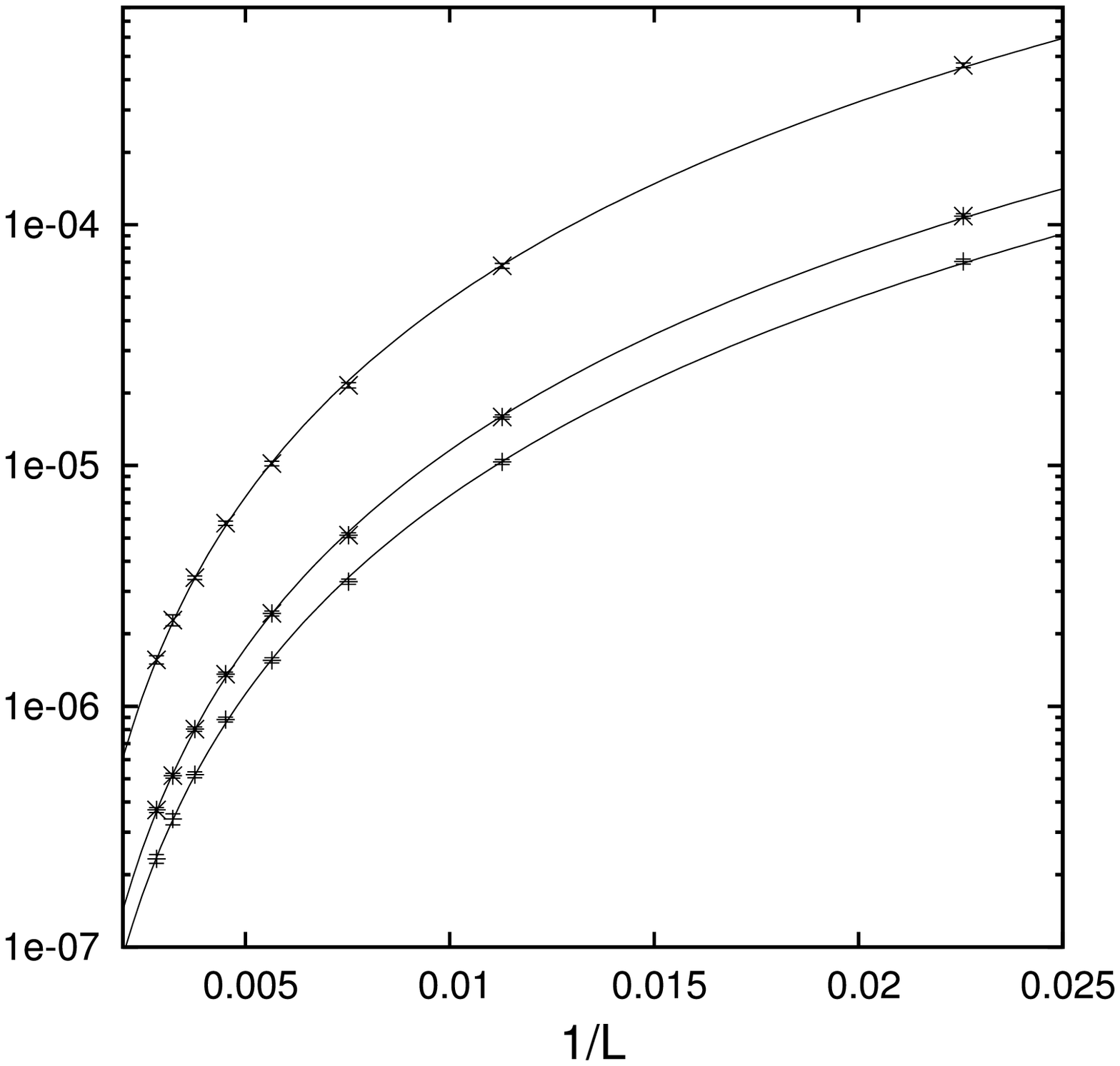}
\vskip 4mm
\includegraphics[scale=0.32]{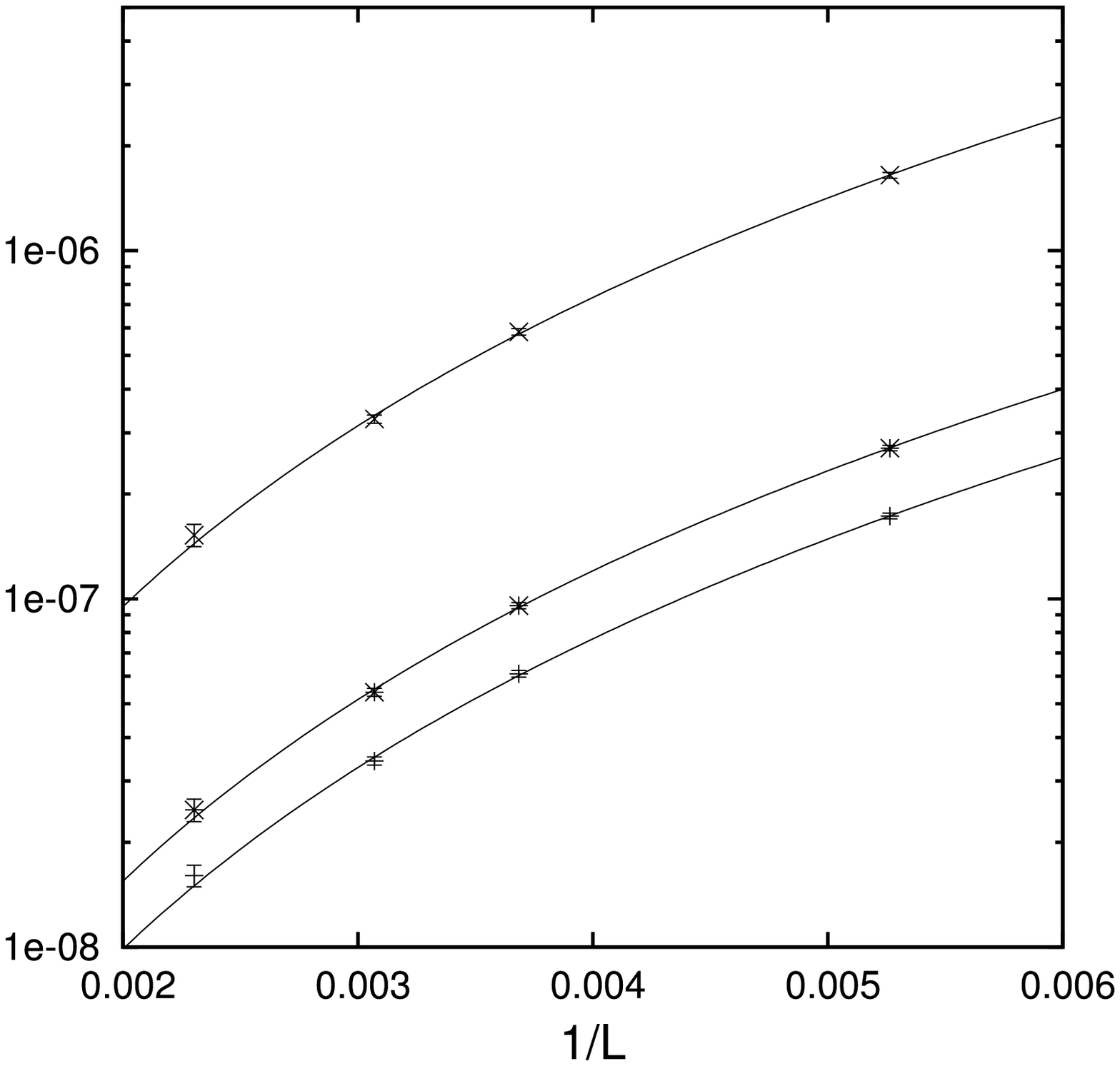}
\vskip 4mm
\includegraphics[scale=0.32]{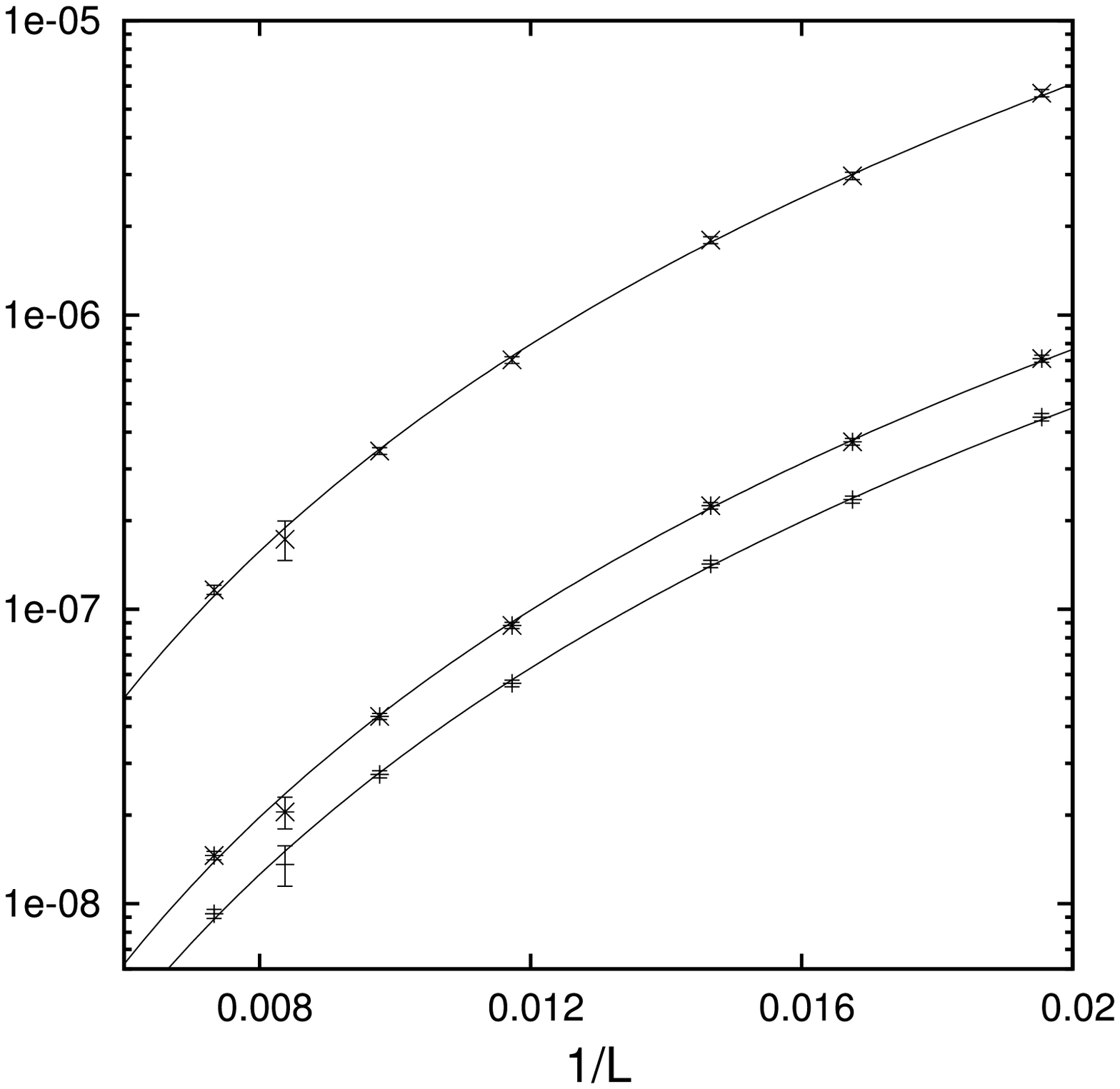}
\caption{\label{fig:A0}
  The square of the quantity $ a \langle {M}(0) \rangle $ and the
  quantity $ a^2 d (N_c^2 - 1) \langle \, {M}(0)^2 \, \rangle $
  (both in GeV$^{-2}$) as a function of the inverse
  lattice side $1/L$ (GeV) for the
  2d case (top), 3d case (center) and the 4d case (bottom). We also show
  the data for $D(0)/V$ (also in GeV$^{-2}$) and
  the fit of the data using the parameters reported in Table \ref{tab:fits}.
  }
\end{figure}

We start by considering the quantity $\langle {M}(0) \rangle$.
We find (see Fig.\ \ref{fig:A0} and Table \ref{tab:fits}) that our data extrapolate very well to
zero as $1 / L^l$, with the values of $l$ given in Table \ref{tab:fits}.
Thus, in 3d and in 4d we have
$\langle {M}(0) \rangle \sim 1 / V^{1/2}$, implying $D(0) > 0$.
In particular, from our fits we obtain $D(0) \geq (B_l / a^l)^2$.
This gives $ D(0) \geq 0.4(1)$
(GeV$^{-2}$) in 3d and $D(0) \geq 2.2(3)$ (GeV$^{-2}$) in 4d,
where we used $a = 1.35687$ GeV$^{-1}$ in 3d and $a = 1.066$ GeV$^{-1}$ in 4d.
As for the upper bound (\ref{eq:ineq2D0}),
using our fits (see again Fig.\ \ref{fig:A0} and Table \ref{tab:fits})
we have $D(0) \leq d(N_c^2 - 1) B_u / a^u$,
yielding $ D(0) \leq 4(1)$ (GeV$^{-2}$) in 3d and
$ D(0) \leq 29(5)$ (GeV$^{-2}$) in 4d. On the other hand, in 2d both the lower and
the upper bounds extrapolate to zero, implying $D(0) = 0$ in agreement with Ref.\
\cite{Maas:2007uv}. Let us note that our bounds in 3d and in 4d are in agreement
with the data shown in Figs.\ 1 and 2 of Ref.\ \cite{Cucchieri:2007md}.
[In the  3d case, compared to the extrapolation reported
in Fig.\ 1 of Ref.\ \cite{Cucchieri:2007md}, one should also include here a
factor $\beta = 3.0$, i.e.\ $1.2(3) \leq D(0) \leq 12(3)$].
Also note that in the three cases one finds $B_u \approx B_l^2$ and $u \approx 2 l$.
Indeed one can check that $ \langle {M}(0) \rangle^2 \lesssim   
\langle \, {M}(0)^2 \, \rangle$, implying that the quantity ${M}(0)$ is
almost the same for all Monte Carlo configurations. More precisely, we 
verified for the three cases that 
$\langle \, {M}(0)^2 \, \rangle - \langle {M}(0) \rangle^2$ [i.e.\ the 
susceptibility of $M(0)$] goes to zero as $\sim 1/V$ in the
infinite-volume limit.

In order to interpret these results, les us first note that,
given a Gaussian random variable $x$ with null mean value
and standard deviation $\sigma$, the random variable $| x |$ has mean value
(and standard deviation) proportional to $\sigma$. In our case, this suggests
that the average value of the gluon field at zero momentum $A(0)$ [defined
in Eq.\ (\ref{eq:A0field})] should be zero with a standard deviation of the order of
$1/L^c$, with $c\approx l$ (see Table \ref{tab:fits}).
This is indeed the case in 2d, 3d and in 4d. [We find respectively
$c= 1.36(2), 1.47(3), 1.97(1)$ for the three cases.] In 3d and in 4d our results
imply $\sigma \propto 1/\sqrt{V}$. This property is known as
{\em self-averaging} \cite{libro} and is the behavior expected for extensive 
quantities in pure phases, away from phase boundaries.
(In our case the magnetization is not extensive because we divide by the
volume, but the result holds for the relative standard deviation.)
More precisely, one talks of strong self-averaging when $\sigma \propto 1/L^c$ 
and $c = d/2$, and of weak self-averaging when $c < d/2$. Thus, we find strong
self-averaging for ${M}(0)$
in 3d and in 4d and some kind of {\em over} self-averaging
in 2d, with $c > d/2$. In simpler terms, the gluon propagator
may be thought of as the susceptibility associated to the magnetization 
$M(0)$ [or rather to the quantity defined by Eq.\ (\ref{eq:Aabs}) {\em without} 
the absolute value, which has zero average]. 
In 3d and 4d the system has (finite) nonzero 
susceptibility, while for 2d the susceptibility is zero.
We do not have a simple explanation for this latter result.
Here we can only argue that the 2d case is probably different since there are no
propagating degrees of freedom.

\begin{table}[t]
\vskip -2mm
\caption{Fits of $ a \langle {M}(0) \rangle $,
         $ a^2 \langle \, {M}(0)^2 \, \rangle $ and $D(0)/V$
         respectively using the Ans\"atze
         $B_l / L^l$, $B_u / L^u$ and $B / L^k$.
         Note that $B_l$, $B_u$ and $B$ have mass dimensions
         respectively $-l-1$, $-u-2$ and $-k-2$.
         Note also that in order to obtain Fig.\ \ref{fig:A0} one should
         multiply by $d (N_c^2 - 1)$ the data and the fit related to 
         the fourth and fifth columns of the table.
         Most of the data used for the fits
         have a statistical error of the order of 2--3 \%.
         For all fits we have $\chi^2/d.o.f. \approx 1$.
\label{tab:fits}}
\begin{tabular}{ccccccc}
 $d$ & $B_l$ & $l$ & $B_u$ & $u$ & $B$ & $k$ \\
\hline
  2  & 1.48(6) & 1.367(8) & 2.3(2)  & 2.72(1)  & 3.3(2) & 2.73(1)  \\
  3  & 1.0(1)  & 1.48(3)  & 1.0(3)  & 2.95(5)  & 1.5(3) & 2.96(4)  \\
  4  & 1.7(1)  & 1.99(2)  & 3.1(5)  & 3.99(4)  & 4.7(8) & 3.99(4)  \\
\hline
\end{tabular}
\end{table}

Note that our results in 3d and in 4d only imply that reflection positivity
is not {\em maximally} violated. A clear violation of 
reflection positivity \cite{violation,prepar}
is still obtained in 2d, 3d and in 4d for the SU(2)
and SU(3) cases.

%%%%%%%%%%%%%%%%%%%%%%%%%%%%%%%%%%%%%%%%%%%%%%%%%%%%%%%%%%%%%%%%%%%%%%%%%%%%%%%%%%%%%%%%

\section{Conclusions}
\label{sec:conclusions}

We have shown that
the Landau-gauge gluon
propagator at zero momentum $D(0)$ is finite and nonzero in 3d and in 4d. At the
same time, we find $D(0) = 0$ in 2d, in agreement with Ref.\ \cite{Maas:2007uv}. These
results have been obtained by considering the inequalities in Eqs.\ (\ref{eq:ineq1D0})
and (\ref{eq:ineq2D0}), i.e.\ by studying the ``magnetization-like'' quantity ${M}(0)$
instead of the ``susceptibility'' $D(0)$. This allows control of the extrapolation
of the data to infinite volume. Moreover, the quantity $D(0)/V$ can be well fitted
in this limit as a function of $1/L$. Our results in 3d and in 4d can be explained
as a manifestation of strong self-averaging.
As mentioned above, a similar analysis may be applied to more general cases,
and considering also nonzero momenta.

%%%%%%%%%%%%%%%%%%%%%%%%%%%%%%%%%%%%%%%%%%%%%%%%%%%%%%%%%%%%%%%%%%%%%%%%%%%%%%%%%%%%%%%%

\section{Acknowledgements}

We thank S. Sorella and D. Zwanziger for helpful discussions.
We acknowledge partial support from FAPESP and from CNPq. The work of T.M. is
supported also by a fellowship from the Alexander von Humboldt Foundation.
Most of the simulations reported here have been done on the IBM
supercomputer at S\~ao Paulo University (FAPESP grant \# 04/08928-3).

%%%%%%%%%%%%%%%%%%%%%%%%%%%%%%%%%%%%%%%%%%%%%%%%%%%%%%%%%%%%%%%%%%%%%%%%%%%%%%%%%%%%%%%%

%%%%%%%%%%%%%%%%%%%%%%%%%%%%%%%%%%%%%%%%%%%%%%%%%%%%%%%%%%%%%%%%%%%%%%%%%%%%%%%%%%%%%%%%

\end{document}